\documentclass[conference]{IEEEtran}
\IEEEoverridecommandlockouts

\usepackage{cite}
\usepackage{amsmath,amssymb,amsfonts}
\usepackage{algorithmic}
\usepackage{graphicx}
\usepackage{textcomp}
\usepackage{xcolor}
\def\BibTeX{{\rm B\kern-.05em{\sc i\kern-.025em b}\kern-.08em
    T\kern-.1667em\lower.7ex\hbox{E}\kern-.125emX}}

\usepackage{booktabs}
\usepackage{hyperref}
\usepackage{subcaption} 
\usepackage{physics}
\usepackage{multirow}

\usepackage{tikz}
\usetikzlibrary{arrows.meta,positioning}

\usepackage{pgfplots}
\pgfplotsset{compat=1.18}

\begin{document}

\title{Failure-Guided Fuzzing for Hybrid Quantum–Classical Programs}

\author{\IEEEauthorblockN{Lei Zhang}
\IEEEauthorblockA{\textit{Department of Information Systems} \\
\textit{University of Maryland, Baltimore County}\\
Maryland, USA \\
leizhang@umbc.edu}
}

\maketitle

\begin{abstract}
Hybrid quantum-classical (HQC) algorithms, such as the Variational Quantum Eigensolver (VQE) and the Quantum Approximate Optimization Algorithm (QAOA), are central to near-term quantum computing but remain challenging to test. Sampling-based fuzzing can expose faulty or non-convergent configurations, but under realistic execution budgets, it may miss failure-prone regions in the joint space of classical optimizer settings and quantum circuit parameters.

This paper studies failure-guided fuzzing for HQC programs. It models a hybrid input as a pair of classical optimizer hyperparameters and quantum circuit parameters, and evaluates a two-phase strategy that first searches for non-convergent seeds and then locally fuzzes circuit parameters around those seeds. To understand where the gains come from, five budgeted strategies are compared: random hybrid testing, classical enumeration without fuzzing, random-seed local fuzzing, enumeration-seed local fuzzing, and concolic-seed local fuzzing. The study is implemented on a VQE instance and a QAOA MaxCut instance in Qiskit. The results show that failure-guided local fuzzing is the main driver of improvement over random testing, while concolic seed discovery provides additional benefits on VQE but is less stable on QAOA. These findings suggest that reusing failure information is a promising direction for HQC testing, but that the value of concolic seed discovery is workload-dependent.
\end{abstract}


\section{Introduction}
\label{sec:introduction}

Hybrid quantum-classical (HQC) algorithms, such as the Variational Quantum Eigensolver (VQE)~\cite{tilly2022variational} and the Quantum
Approximate Optimization Algorithm (QAOA)~\cite{fakhimi2023quantum},
are a central pillar of the current Noisy Intermediate-Scale Quantum
(NISQ) era~\cite{bharti2022noisy}. In these workloads, a classical optimizer iteratively updates circuit parameters, invokes a parameterized quantum circuit on a backend or simulator, and uses measurement outcomes to drive subsequent updates. This tight feedback
loop between classical control logic and quantum execution makes HQC programs difficult to test: failures may arise from the quantum circuit, the optimizer, their interaction, or the stochastic nature of
measurements~\cite{zhang2023identifying, kim2025detecting}.

Recent work in quantum software engineering has begun to adapt techniques such as unit testing~\cite{miranskyy2025feasibility,
miranskyy2019testing}, mutation testing~\cite{fortunato2022qmutpy,
mendiluze2021muskit}, and fuzzing~\cite{wang2018quanfuzz} to the quantum setting. In particular, sampling-based fuzzing has been proposed as a way to explore large quantum input spaces by generating many random inputs and searching for crash-inducing executions~\cite{wang2018quanfuzz}. While such techniques can be effective on small examples, they face a fundamental rare-event barrier: under realistic testing budgets, random sampling may fail to hit combinations of classical hyperparameters and quantum circuit parameters that trigger non-convergence or other faulty behavior.

Moreover, many existing quantum testing approaches either treat the classical and quantum components largely in isolation, or reason about quantum behavior at the level of circuits and states without explicitly leveraging the structure of the classical control flow and optimizer hyperparameters~\cite{zhao2020quantum}. This leaves a gap: testing methods are needed that can systematically explore the classical control side of HQC programs and then focus fuzzing effort on regions of the circuit-parameter space that appear failure-prone.

This paper studies \emph{failure-guided fuzzing} for HQC programs. A hybrid input is modeled as a pair of hybrid inputs $h = (c,\boldsymbol{\theta})$, where $c$ denotes classical hyperparameters (e.g., optimizer type, iteration limit, and learning rate) and $\boldsymbol{\theta}$ denotes quantum circuit parameters (e.g., rotation angles in a VQE or QAOA ansatz). A run is classified as a \emph{crash} if the hybrid execution fails to converge to a workload-specific target value within a fixed resource budget; this serves as a non-convergence oracle.

The prototype evaluates a two-phase failure-guided strategy. First, a seed-discovery phase searches for non-convergent hybrid inputs using random sampling, classical enumeration, or lightweight concolic exploration over discrete classical hyperparameters. In the structured two-phase variants, ENUM-FUZZ uses enumerative seed discovery followed by local fuzzing, while SYM-FUZZ uses an SMT-backed (Satisfiability Modulo Theories~\cite{de2008z3}) symbolic explorer to obtain path-diverse classical inputs~\cite{sen2007concolic}. Second, the local fuzzing phase mutates circuit parameters around crash-inducing seeds while keeping the classical configuration fixed~\cite{godefroid2008automated}.

The study is instantiated on two small but representative HQC workloads: a 2-qubit VQE instance with a simple variational Hamiltonian and a 4-qubit QAOA MaxCut instance on a ring graph, both implemented in Qiskit~\cite{javadi2024quantum}.\footnote{The prototype targets small VQE and QAOA instances on noiseless simulators. The goal is not to claim scalability to large real-world workloads, but to assess how different budgeted testing strategies behave on representative HQC patterns.} To disentangle the sources of testing effectiveness, five strategies are compared under the same execution budget. The first two are non-fuzzing baselines: random hybrid testing (RH), which samples both $c$ and $\boldsymbol{\theta}$ from fixed global distributions, and ENUM, which performs classical enumeration without local fuzzing. The remaining three are failure-guided local-fuzzing strategies: RAND-FUZZ, which fuzzes around the worst crash seeds found by random testing; ENUM-FUZZ, which fuzzes around crash seeds found by classical enumeration; and SYM-FUZZ, which fuzzes around crash seeds found by concolic seed discovery. Together, these strategies separate the effects of unguided random exploration, classical exploration, local fuzzing, and concolic seed discovery.

This work makes 3 \textbf{contributions}. First, it formulates failure-guided fuzzing for HQC programs by treating $(c,\boldsymbol{\theta})$ as the testing object and using non-convergence as a budget-aware oracle; it then introduces a two-phase prototype that combines crash-seed discovery with local circuit-parameter fuzzing. Second, it empirically disentangles the roles of random exploration, classical enumeration, local fuzzing, and concolic seed discovery through a five-strategy evaluation on two case studies, i.e., VQE and QAOA. Last but not least, the artifacts, including source code and experimental results, are publicly available on Zenodo at \href{https://doi.org/10.5281/zenodo.19718515}{doi:10.5281/zenodo.19718515}.

\section{Motivation: Rare-Event Challenges}
\label{sec:background}

\begin{figure}[t]
  \centering
  \begin{tikzpicture}
    \begin{axis}[
      width=0.9\columnwidth,
      height=0.55\columnwidth,
      ymode=log,
      xlabel={Number of qubits $N_q$},
      ylabel={Expected crashes $B\cdot 2^{-N_q}$},
      xmin=2, xmax=24,
      ymin=1e-3, ymax=1e4,
      grid=both,
      legend style={font=\scriptsize, at={(0.98,0.98)}, anchor=north east},
      tick label style={font=\scriptsize},
      label style={font=\scriptsize},
    ]
      \addplot+[mark=none, thick] table[row sep=\\] {
      x y\\
      2 250\\ 4 62.5\\ 6 15.625\\ 8 3.90625\\ 10 0.97656\\
      12 0.24414\\ 14 0.06104\\ 16 0.01526\\ 18 0.00381\\
      20 0.00095\\ 22 0.00024\\ 24 0.00006\\
      };
      \addlegendentry{$B=10^3$}

      \addplot+[mark=none, thick, dashed] table[row sep=\\] {
      x y\\
      2 2500\\ 4 625\\ 6 156.25\\ 8 39.0625\\ 10 9.7656\\
      12 2.4414\\ 14 0.6104\\ 16 0.1526\\ 18 0.0381\\
      20 0.0095\\ 22 0.0024\\ 24 0.0006\\
      };
      \addlegendentry{$B=10^4$}

      \addplot+[mark=none, thick, dotted] table[row sep=\\] {
      x y\\
      2 25000\\ 4 6250\\ 6 1562.5\\ 8 390.625\\ 10 97.656\\
      12 24.414\\ 14 6.104\\ 16 1.526\\ 18 0.381\\
      20 0.095\\ 22 0.024\\ 24 0.006\\
      };
      \addlegendentry{$B=10^5$}

      \addplot+[mark=none, black] coordinates {(2,1) (24,1)};
      \addlegendentry{$\mathbb{E}[X]=1$}
    \end{axis}
  \end{tikzpicture}
  \caption{Illustration of the rare-event barrier. If a crash occurs only when a specific $N_q$-qubit bitstring is observed and the measurement distribution is close to uniform, then the expected number of crashes under budget $B$ is at most $B\cdot 2^{-N_q}$. Even large fixed budgets quickly fall below the expected crash as $N_q$ grows.}
  \label{fig:rare-event}
\end{figure}
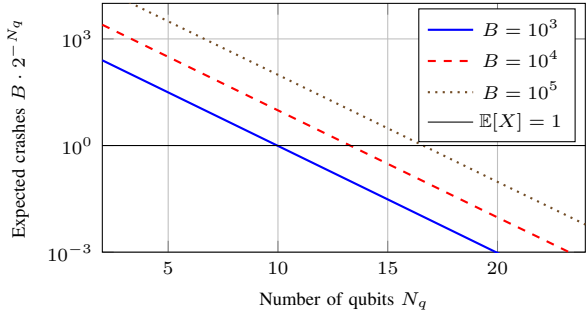

Sampling-based fuzzing has been proposed as a way to explore large quantum input spaces by generating many random inputs and observing the resulting measurement distributions~\cite{wang2018quanfuzz}. However, purely random exploration can face a \emph{rare-event problem}. Consider a toy hybrid program that crashes only when the measured state equals a single $N_q$-qubit bitstring $z^\star$. Let $p_{N_q}$ denote the crash probability of one run:
\[
  p_{N_q} \;=\; \Pr\bigl[\text{measurement outcome} = z^\star\bigr].
\]
If the measurement distribution is close to uniform, then
\[
  p_{N_q} \;\le\; 2^{-N_q}.
\]

Suppose a fuzzing campaign executes the program $B$ times independently. Let $X_{N_q}$ be the random variable counting how many crashes are observed in those $B$ runs. Under the standard Bernoulli model, the expected number of crashes is:
\[
  \mathbb{E}[X_{N_q}] = B \, p_{N_q}
  \;\le\; B \cdot 2^{-N_q}
  \;\xrightarrow[N_q \to \infty]{}\; 0.
\]
Thus, for any fixed budget $B$, the expected number of observed crashes decreases exponentially in $N_q$. This toy example does not model all HQC failures, but it illustrates why unguided sampling may waste substantial budget before reaching failure-prone regions.

Figure~\ref{fig:rare-event} illustrates this scaling behavior for several fixed testing budgets. Even when the budget increases from $10^3$ to $10^5$ executions, the expected number of observed crashes drops rapidly as the number of qubits $N_q$ grows.

This motivates testing strategies that reuse information from observed failures. In this paper, failure-guided local fuzzing reallocates budget around crash-inducing seeds instead of repeatedly sampling fresh hybrid inputs from a global distribution. Concolic guidance is then studied as one way to improve seed discovery by using structural information from the classical optimizer configuration space.

\section{Failure-Guided Fuzzing Approach}
\label{sec:framework}

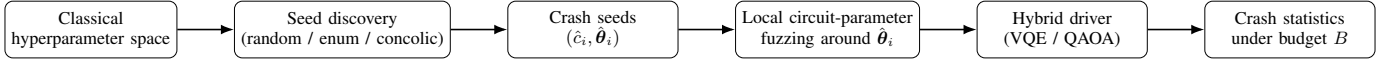
\begin{figure*}[th!]
  \centering
  \resizebox{\textwidth}{!}{%
  \begin{tikzpicture}[
    >=Latex,
    node distance=8mm and 10mm,
    block/.style={
      rectangle,
      draw,
      rounded corners,
      align=center,
      minimum width=30mm,
      minimum height=10mm,
      font=\small
    },
    line/.style={->, thick},
    every node/.style={font=\small}
  ]

    \node[block] (searchspace) {Classical\\hyperparameter space};

    \node[block, right=of searchspace] (seeddisc) {Seed discovery\\(random / enum / concolic)};

    \node[block, right=of seeddisc] (seeds) {Crash seeds\\$(\hat c_i,\hat{\boldsymbol{\theta}}_i)$};

    \node[block, right=of seeds] (fuzz) {Local circuit-parameter\\fuzzing around $\hat{\boldsymbol{\theta}}_i$};

    \node[block, right=of fuzz] (driver) {Hybrid driver\\(VQE / QAOA)};

    \node[block, right=of driver] (stats) {Crash statistics\\under budget $B$};

    \draw[line] (searchspace) -- (seeddisc);
    \draw[line] (seeddisc) -- (seeds);
    \draw[line] (seeds) -- (fuzz);
    \draw[line] (fuzz) -- (driver);
    \draw[line] (driver) -- (stats);

  \end{tikzpicture}%
  }
  \caption{Failure-guided fuzzing workflow: crash seeds are discovered through random, enumerative, or concolic exploration, then used for local circuit-parameter fuzzing under a fixed budget.}
  \label{fig:cf-workflow-compact}
\end{figure*}

This section describes the failure-guided fuzzing prototype. The key idea is to identify crash-inducing hybrid inputs early and reuse them by locally fuzzing nearby circuit parameters. The study evaluates multiple seed-discovery mechanisms---random sampling, classical enumeration, and concolic guidance---to separate the effects of random exploration, structured classical exploration, local fuzzing, and concolic seed discovery. Figure~\ref{fig:cf-workflow-compact} summarizes the workflow.

\subsection{Hybrid Inputs and Failure Oracle}
\label{subsec:oracle}

An HQC program is modeled as a hybrid driver
\[
  \textsc{HybridProg}(c,\boldsymbol{\theta}) \;\to\; \text{trace},
\]
where $c$ denotes classical hyperparameters and
$\boldsymbol{\theta}$ denotes quantum circuit parameters, typically rotation angles in a parameterized ansatz. In the prototype, the classical configuration is
\[
  c = (\textit{opt}, \textit{max\_iter}, \textit{lr}) \in \mathcal{C},
\]
where \textit{opt} is the optimizer type, \textit{max\_iter} is the iteration limit, and \textit{lr} is the learning rate. The circuit parameter vector satisfies $\boldsymbol{\theta}\in\Theta\subseteq\mathbb{R}^d$, with angles identified modulo $2\pi$.

Executing \textsc{HybridProg} on a hybrid input $h$ produces a trace containing classical optimizer states, quantum objective estimates obtained from sampled measurements or statevector simulation, and estimates of a scalar objective $J(h)$, such as the energy in VQE or the MaxCut objective in QAOA. Following common practice in variational algorithm debugging~\cite{hao2023enabling},
the prototype uses \emph{non-convergence} as the test oracle. Given a reference target value $J^\star$ and tolerance $\tau>0$, a run is classified as non-crashing if the best observed cost satisfies $J(h)\leq J^\star+\tau$ and as crashing otherwise. Thus, a crash indicates that the optimizer terminated without reaching the target objective within the allotted resource budget.

\subsection{Overview of the Failure-Guided Strategy}

The prototype separates testing into two responsibilities: discovering crash-inducing hybrid inputs and exploiting them through local fuzzing. Seed discovery can be instantiated by random sampling, classical enumeration, or concolic exploration over discrete classical hyperparameters. Given crash seeds
$(\hat{c},\hat{\boldsymbol{\theta}})$, the fuzzing phase fixes $\hat{c}$ and locally perturbs $\hat{\boldsymbol{\theta}}$ to search nearby failure-prone regions. This design enables the experiments to isolate the effects of classical exploration, local fuzzing, and concolic seed discovery.

\subsection{Seed Discovery over Classical Hyperparameters}
\label{subsec:concolic}

The seed-discovery phase explores the discrete classical hyperparameter space $\mathcal{C}$. It does not symbolically encode quantum states or real-valued circuit parameters; each explored configuration is paired with an initial $\boldsymbol{\theta}$ drawn from the same fixed distribution across strategies.

\textsc{RAND-FUZZ} uses the same random hybrid sampling procedure as \textsc{RH} during seed discovery, then selects crash-inducing inputs for local fuzzing. This strategy isolates the effect of local fuzzing when seeds are obtained without structured classical exploration.

\textsc{ENUM-FUZZ} uses enumeration to explore all configurations in $\mathcal{C}$ and runs \textsc{HybridProg} once for each pair $(c,\boldsymbol{\theta})$. Used without the fuzzing phase, the same enumerative exploration gives the \textsc{ENUM} baseline, which measures the value of structured classical exploration alone. When crash-inducing inputs found by enumeration are used as seeds for local fuzzing, the resulting two-phase strategy is \textsc{ENUM-FUZZ}.

\textsc{SYM-FUZZ} uses a lightweight SMT-backed concolic engine during seed discovery. The engine introduces integer indices representing optimizer type, iteration limit, and learning rate. During execution, branch
predicates over these discrete variables are recorded and selectively flipped to obtain path-diverse classical configurations~\cite{sen2007concolic,de2008z3}. The search maintains a worklist of path states and stops when the worklist is empty or a small bound $m_{\max}$ is reached. Each explored configuration is executed
with an initial $\boldsymbol{\theta}$, producing a result
$r=(c,\boldsymbol{\theta},\mathsf{crash}(h),J(h))$.

\subsection{Crash Seed Selection}
\label{subsec:seed-selection}

Given seed-discovery results $R_{\mathit{seed}}$, the prototype filters to crashing runs with defined costs and ranks them by severity. Severity is measured by how far the best observed objective remains from the target after normalizing the objective to a minimization form. The prototype then greedily selects up to $k$ seeds, prioritizing the most severe crashes while avoiding duplicate classical configurations $c = (\mathit{opt},\mathit{max\_iter},\mathit{lr})$. This diversity constraint prevents the fuzzing phase from spending the remaining budget on multiple seeds with the same classical setting.

\subsection{Local Fuzzing of Circuit Parameters}
\label{subsec:fuzzing}

Given a crash seed $(\hat{c},\hat{\boldsymbol{\theta}})$, the fuzzing phase explores nearby circuit-parameter configurations that may also induce non-convergence. The classical configuration $\hat{c}$ is kept fixed, and each new parameter vector is sampled as
\[
  \boldsymbol{\theta}' =
  \hat{\boldsymbol{\theta}} + \boldsymbol{\eta}
  \pmod{2\pi},
  \qquad
  \boldsymbol{\eta} \sim \mathcal{N}(0,\sigma^2 I).
\]
The scale $\sigma$ controls locality: smaller values stay close to the seed, while larger values explore more diverse but less tightly localized inputs.

Given a fuzzing budget $B_{\mathit{fuzz}}$, the prototype distributes the budget evenly across the selected seeds, allocating approximately $B_{\mathit{fuzz}}/k$ runs per seed. Each fuzz run executes \textsc{HybridProg} on $(\hat{c},\boldsymbol{\theta}')$ using the same backend and measurement settings as in seed discovery, then applies the non-convergence oracle. The resulting crash and non-crash outcomes are aggregated to compute the crash counts and rates for the fuzzing phase.

This design assumes that non-convergent executions often occur in local neighborhoods of the variational parameter space. This assumption is plausible for variational algorithms with relatively smooth cost landscapes, but may be weaker for more rugged workloads such as QAOA.

\section{Experimental Design}
\label{sec:experimental-design}

This section describes the experimental methodology for evaluating failure-guided fuzzing on the VQE and QAOA case studies. The evaluation is driven by three research questions:

\textbf{RQ1.}
Under a fixed budget, do local-fuzzing strategies discover more non-convergence failures than RH and ENUM?

\textbf{RQ2.}
Does concolic seed discovery in SYM-FUZZ improve over random-seed local fuzzing, and is the effect workload-dependent?

\textbf{RQ3.}
How stable are the strategies across repeated trials and across the two workloads?

To answer these questions, the experiments compare five strategies under identical execution budgets: RH, ENUM, ENUM-FUZZ, RAND-FUZZ, and SYM-FUZZ.

\subsection{Experimental Setup}
\label{subsec:env}

All experiments are implemented in Python using Qiskit and the Aer simulator as the quantum backend~\cite{chundury2025quantum}. A budgeted execution model is used to ensure a fair comparison. For each workload (VQE or QAOA) and each strategy, the experiment runs $T=20$ independent trials, each under the same budget $B=2{,}000$ program runs but with different random seeds. A program run denotes one call to the hybrid driver with a specific hybrid input $h$, including all quantum circuit executions required by the optimizer.

Each trial records the total number of crashes $C$ and the crash rate $C/B$. Results are reported as the mean and standard deviation of crash counts $C$ and crash rates $C/B$ over the $T$ trials. The five strategies are:

\begin{enumerate}
  \item \textbf{RH}: random hybrid testing, where each run samples a fresh classical configuration $c$ and circuit-parameter vector $\boldsymbol{\theta}$ from fixed global distributions.
  \item \textbf{ENUM}: classical enumeration without local fuzzing, which isolates the effect of structured exploration of $\mathcal{C}$ alone.
  \item \textbf{ENUM-FUZZ}: local fuzzing around crash seeds found by classical enumeration.
  \item \textbf{RAND-FUZZ}: local fuzzing around the worst crash seeds found by random hybrid testing.
  \item \textbf{SYM-FUZZ}: local fuzzing around crash seeds found by a lightweight concolic/symbolic explorer.
\end{enumerate}

For the local-fuzzing strategies (RAND-FUZZ, ENUM-FUZZ, and SYM-FUZZ), the seed-selection rule, fuzzing scale, and fuzzing budget are kept fixed for a given workload. Thus, differences among these strategies reflect how crash seeds are discovered rather than changes in the local fuzzing procedure.

\subsection{Workload-Specific Configuration}
\label{subsec:config}

\begin{table}[t]
\centering
\caption{Workload and testing parameters used in the main experiments.}
\label{tab:workload-config}
\resizebox{\columnwidth}{!}{%
\begin{tabular}{lll}
\toprule
Item & VQE & QAOA \\
\midrule
Qubits & 2 & 4 \\
Circuit & 2-layer $R_y$ ansatz & depth-$P$ ring MaxCut ($P=2$) \\
Classical params & opt, max\_iter, lr & opt, max\_iter, lr \\
Iterations & $\{5,10,20\}$ & $\{5,10,20\}$ \\
Budget $B$ & 2,000 & 2,000 \\
Trials $T$ & 20 & 20 \\
Oracle & non-convergence & non-convergence \\
Tolerance & 0.3 & 0.5 \\
$\sigma$ & 0.2 & 0.2 \\
Shots & N/A (statevector) & 1024 \\
\bottomrule
\end{tabular}%
}
\end{table}

Table~\ref{tab:workload-config} summarizes the workload and testing parameters used in the main experiments. The budgeted design and strategy definitions are kept identical for VQE and QAOA; only the hybrid driver and workload-specific oracle differ. Both workloads use the same structure of classical hyperparameters: optimizer type, maximum iteration count, and learning rate.

For VQE, the objective is normalized so that lower values indicate better energy estimates, and non-convergence is defined as failing to reach $J^\star+\tau$ within the iteration budget. For QAOA, the MaxCut objective is converted to the same minimization form by using the gap between the maximum cut value and the observed cut value. For the 4-node ring graph, the maximum cut value is 4.

The non-convergence tolerance and local fuzzing scale $\sigma$ are fixed in the main experiments. These settings are chosen so that RH observes a non-trivial but not overwhelming crash rate, allowing the experiments to compare how effectively each strategy uses the same budget. Section~\ref{sec:results} also reports a sensitivity analysis over the tolerance and $\sigma$ values.

\section{Experimental Results}
\label{sec:results}

This section reports the results of the five-strategy evaluation on the VQE and QAOA workloads. The analysis focuses on two budget-aware metrics: crash count and crash rate. Crash count measures how many non-convergent executions are observed within the fixed budget, while crash rate normalizes this count by the total number of hybrid driver calls.

These metrics indicate how effectively each strategy reaches failure-prone regions of the hybrid input space. They do not necessarily represent distinct root-cause faults, since nearby crashes may reflect multiple manifestations of the same underlying issue.

\subsection{VQE Case Study}
\label{subsec:results-vqe}

\begin{figure}[t]
  \centering
  \includegraphics[width=0.45\linewidth]{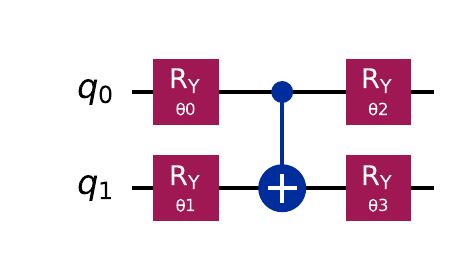}
  \caption{Parametrized 2-qubit VQE ansatz used in the case study: two layers of $R_y$ rotations separated by a CNOT entangling gate.}
  \label{fig:vqe-ansatz}
\end{figure}

\begin{table}[t]
  \centering
  \caption{VQE: crash statistics over $T=20$ trials with budget $B=2{,}000$ hybrid runs per trial. Values are reported as mean $\pm$ standard deviation.}
  \label{tab:vqe-crash-stats}
  \begin{tabular}{lcc}
    \toprule
    Strategy & Crashes per trial & Crash rate \\
    \midrule
    RH 
      & $200.9 \pm 17.8$ 
      & $0.100 \pm 0.009$ \\
    ENUM
      & $494.6 \pm 9.6$
      & $0.247 \pm 0.005$ \\
    ENUM-FUZZ 
      & $1387.3 \pm 223.7$
      & $0.694 \pm 0.112$ \\
    RAND-FUZZ
      & $1430.5 \pm 86.9$
      & $0.715 \pm 0.043$ \\
    SYM-FUZZ 
      & $1513.8 \pm 66.3$
      & $0.757 \pm 0.033$ \\
    \bottomrule
  \end{tabular}
\end{table}

\begin{figure}[t]
  \centering
  \includegraphics[width=0.7\linewidth]{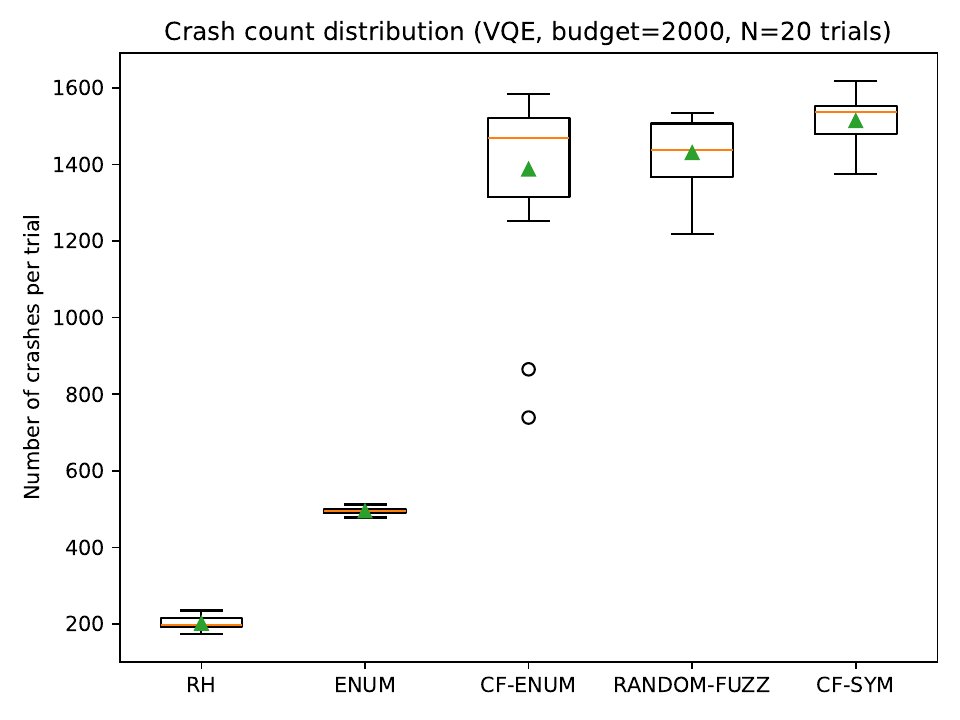}
  \caption{VQE: distribution of crash counts per trial for each strategy over $T=20$ trials with budget $B=2{,}000$. Boxes show interquartile ranges; whiskers and markers indicate spread and outliers.}
  \label{fig:vqe-boxplot}
\end{figure}


The VQE case study follows the standard variational structure: a parameterized ansatz $U(\boldsymbol{\theta})$ prepares a quantum state $\ket{\psi(\boldsymbol{\theta})}$, measurements estimate the objective associated with a problem Hamiltonian, and a classical optimizer updates the circuit parameters. The prototype uses a 2-qubit hardware-efficient ansatz with two layers of $R_y$ rotations separated by a CNOT gate, as shown in Fig.~\ref{fig:vqe-ansatz}.

\textbf{Crash counts and crash rates.}
Table~\ref{tab:vqe-crash-stats} and Fig.~\ref{fig:vqe-boxplot} show that all structured or failure-guided strategies outperform RH on VQE. RH finds $200.9 \pm 17.8$ crashes per trial, corresponding to a crash rate of $0.100 \pm 0.009$. ENUM improves this to $494.6 \pm 9.6$ crashes per trial, indicating that structured exploration of the classical hyperparameter space alone is useful, but still limited.

The largest gains come from strategies that reuse crash information through local fuzzing. The three failure-guided fuzzing strategies, i.e., ENUM-FUZZ, RAND-FUZZ, and SYM-FUZZ find $1387.3 \pm 223.7$, $1430.5 \pm 86.9$, and $1513.8 \pm 66.3$ crashes per trial, respectively. Thus, local fuzzing around failure-prone seeds is the dominant source of improvement over RH and ENUM. Among the local-fuzzing strategies, SYM-FUZZ performs best, achieving the highest mean crash count and the
lowest variance among the three high-performing methods. This suggests that, for the VQE workload, lightweight concolic seed discovery can improve the quality of seeds supplied to the local fuzzing phase.

Overall, the VQE results support a layered interpretation: classical enumeration improves over random testing, local fuzzing around crash seeds yields a much larger gain, and concolic seed discovery appears to provide an additional benefit when used to select seeds for local fuzzing. The exported crash seeds further suggest that severe VQE failures often combine unfavorable optimizer settings with circuit-parameter vectors that fail to reach the target energy within the allotted optimization budget.

\subsection{QAOA Case Study}
\label{subsec:results-qaoa}

\begin{figure}[t]
  \centering
  \includegraphics[width=0.95\linewidth]{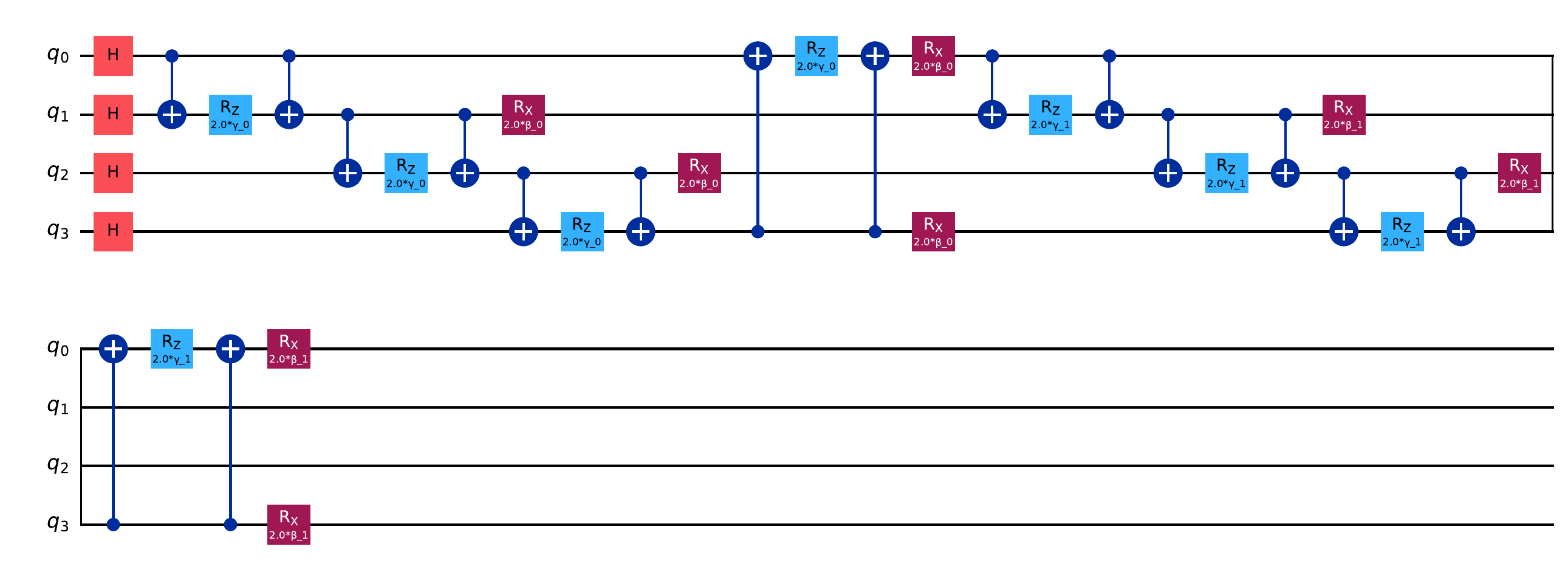}
  \caption{QAOA ansatz used in the case study: depth-$P$ QAOA for MaxCut on a 4-node ring graph, starting from $\ket{+}^{\otimes n}$ and alternating cost layers with $ZZ(\gamma_\ell)$ rotations on each edge and mixer layers with $RX(2\beta_\ell)$ rotations on all qubits.}
  \label{fig:qaoa-ansatz}
\end{figure}

\begin{table}[t]
  \centering
  \caption{QAOA: crash statistics over $T=20$ trials with budget $B=2{,}000$ hybrid runs per trial. Values are reported as mean $\pm$ standard deviation.}
  \label{tab:qaoa-crash-stats}
  \begin{tabular}{lcc}
    \toprule
    Strategy & Crashes per trial & Crash rate \\
    \midrule
    RH
      & $305.9 \pm 13.5$
      & $0.153 \pm 0.007$ \\
    ENUM
      & $301.8 \pm 13.1$
      & $0.151 \pm 0.007$ \\
    ENUM-FUZZ
      & $721.4 \pm 329.5$
      & $0.361 \pm 0.165$ \\
    RAND-FUZZ
      & $861.7 \pm 119.5$
      & $0.431 \pm 0.060$ \\
    SYM-FUZZ
      & $773.6 \pm 427.6$
      & $0.387 \pm 0.214$ \\
    \bottomrule
  \end{tabular}
\end{table}

\begin{figure}[t]
  \centering
  \includegraphics[width=0.7\linewidth]{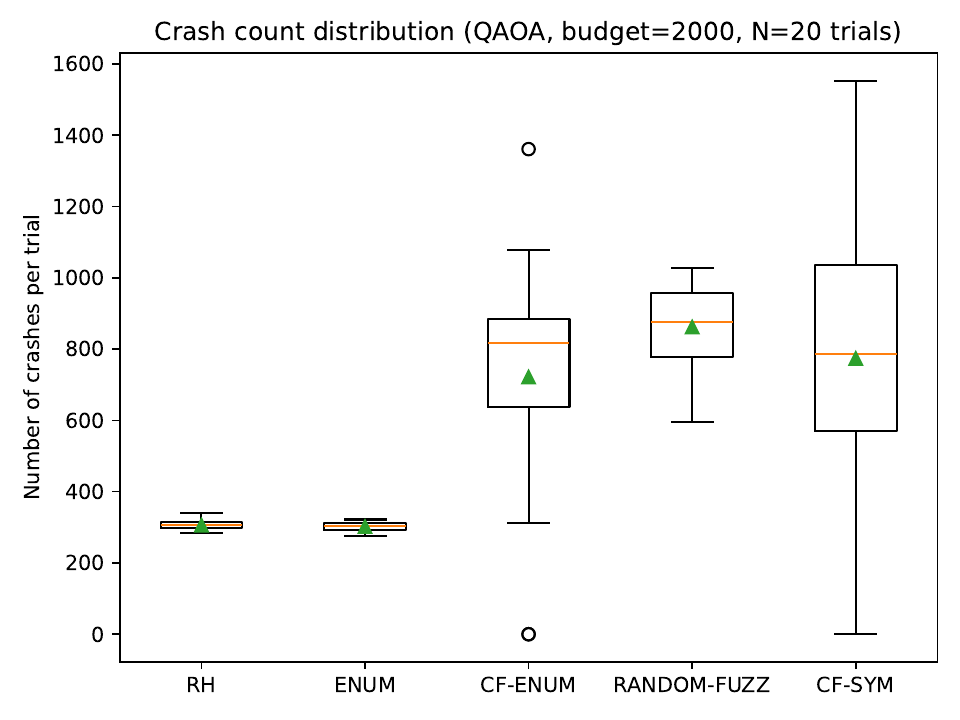}
  \caption{QAOA: distribution of crash counts per trial for each strategy over $T=20$ trials with budget $B=2{,}000$.}
  \label{fig:qaoa-boxplot}
\end{figure}


The second case study is a QAOA instance for MaxCut on a 4-node ring graph. Figure~\ref{fig:qaoa-ansatz} shows the $P$-layer QAOA ansatz: the circuit prepares $\ket{+}^{\otimes n}$ and then alternates cost layers, implemented by two-qubit $ZZ(\gamma_\ell)$ rotations along the graph edges, with mixer layers, implemented by single-qubit $RX(2\beta_\ell)$ rotations on all qubits.

\textbf{Crash counts and crash rates.}
Table~\ref{tab:qaoa-crash-stats} and Fig.~\ref{fig:qaoa-boxplot} show that QAOA has a higher RH baseline crash rate than VQE. RH finds $305.9 \pm 13.5$ crashes per trial, corresponding to a crash rate of $0.153 \pm 0.007$. ENUM performs almost identically to RH, with $301.8 \pm 13.1$ crashes per trial and a crash rate of $0.151 \pm 0.007$. This suggests that classical enumeration alone does not provide a meaningful advantage for this workload.

As in VQE, the largest gains come from strategies that reuse crash information through local fuzzing. RAND-FUZZ achieves the highest mean crash count on QAOA, with $861.7 \pm 119.5$ crashes per trial and a crash rate of $0.431 \pm 0.060$. ENUM-FUZZ and SYM-FUZZ also outperform RH and ENUM, reaching $721.4 \pm 329.5$ and $773.6 \pm 427.6$ crashes per trial, respectively. However, unlike in the VQE case, neither concolic-guided strategy clearly exceeds RAND-FUZZ on average, and SYM-FUZZ exhibits particularly high variance.

These results indicate that local fuzzing remains highly effective on QAOA, but the added value of concolic seed discovery is weaker and less stable than in VQE. A plausible explanation is that the QAOA failure landscape is more rugged or fragmented: random seed discovery can already find failure-prone regions, while symbolic exploration over the small discrete classical space does not consistently identify better seeds. Representative problematic QAOA seeds often combine small iteration budgets or aggressive learning rates with circuit parameter vectors that remain far from the target cut value after optimization.

\subsection{Statistical and Robustness Checks}
\label{subsec:additional-analyses}

\begin{table}[t]
\centering
\caption{Selected Mann--Whitney U test~\cite{mann1947test} is employed for pairwise
comparisons of per-trial crash counts, and Cliff's
$\delta$~\cite{cliff1993dominance} is used to report effect size.}
\label{tab:significance}
\resizebox{\columnwidth}{!}{%
\begin{tabular}{llrr}
\toprule
Workload & Comparison & $p$-value & Cliff's $\delta$ \\
\midrule
VQE  & RH vs. ENUM         & $5.99\times10^{-8}$ & $1.00$ \\
VQE  & RH vs. RAND-FUZZ  & $6.22\times10^{-8}$ & $-1.00$ \\
VQE  & RH vs. ENUM-FUZZ      & $6.23\times10^{-8}$ & $-1.00$ \\
VQE  & RH vs. SYM-FUZZ       & $6.23\times10^{-8}$ & $-1.00$ \\
VQE  & SYM-FUZZ vs. RAND-FUZZ & $6.82\times10^{-3}$ & $0.50$ \\
\midrule
QAOA & RH vs. ENUM         & $4.65\times10^{-1}$ & $0.14$ \\
QAOA & RH vs. RAND-FUZZ  & $6.27\times10^{-8}$ & $-1.00$ \\
QAOA & RH vs. ENUM-FUZZ      & $3.28\times10^{-5}$ & $-0.77$ \\
QAOA & RH vs. SYM-FUZZ       & $1.52\times10^{-4}$ & $-0.70$ \\
QAOA & SYM-FUZZ vs. RAND-FUZZ & $7.05\times10^{-1}$ & $-0.07$ \\
\bottomrule
\end{tabular}%
}
\end{table}

Table~\ref{tab:significance} supports the main trends. On VQE, ENUM-FUZZ, RAND-FUZZ, and SYM-FUZZ all significantly outperform RH, and SYM-FUZZ also outperforms RAND-FUZZ, indicating that concolic seed discovery adds benefit beyond unguided local fuzzing on this workload. On QAOA, all three local-fuzzing strategies significantly outperform RH, whereas ENUM does not; SYM-FUZZ is also not significantly different from RAND-FUZZ. These results support the conclusion that local fuzzing is robustly useful, while the added value of concolic guidance is workload-dependent.

A sensitivity check varied the non-convergence tolerance and local fuzzing scale $\sigma$. The main trends remain stable across all tested settings: all local-fuzzing strategies continued to outperform RH and ENUM. On VQE, SYM-FUZZ is strongest in most settings and achieves the highest peak mean crash count, ranging from $945.5$ to $1832.7$ mean crashes, while RAND-FUZZ led in some looser threshold or larger-$\sigma$ settings. On QAOA, RAND-FUZZ is strongest or near-strongest across settings, with mean crash counts ranging from $658.2$ to $1159.1$. Detailed per-trial results, exported seeds, clustering summaries, and scripts are included in the artifact.

\subsection{Answers to RQs}

\textbf{RQ1.}
Across both workloads, strategies that reuse crash information through local fuzzing substantially outperform RH and ENUM under the same budget. This shows that the main gain comes from local circuit-parameter fuzzing around failure-prone seeds, rather than from classical exploration.

\textbf{RQ2.}
The benefit of concolic seed discovery is workload-dependent. On VQE, SYM-FUZZ performs best, suggesting that symbolic exploration can identify higher-quality seeds for local fuzzing. On QAOA, RAND-FUZZ matches or exceeds ENUM-FUZZ and SYM-FUZZ on average, indicating that random seed discovery plus local fuzzing can be competitive on more fragmented failure landscapes.

\textbf{RQ3.}
The strategies are more stable on VQE than on QAOA. SYM-FUZZ achieves the highest mean crash count with relatively low variance on VQE, whereas the QAOA results show larger run-to-run variability, especially for ENUM-FUZZ and SYM-FUZZ. Overall, failure-guided local fuzzing is robustly useful, but the added value of concolic guidance depends on the workload.

\section{Threats to Validity}
\label{subsec:threats}

\textbf{Internal validity.}
The hybrid drivers, seed-discovery procedures, and fuzzing logic are implemented in-house, so implementation bugs could bias the results. This risk is mitigated by validating the non-convergence oracles on hand-checked configurations, cross-checking results with simpler baselines, and inspecting crash distributions and exported seed sets.

\textbf{Construct validity.}
Crash count and crash rate are budget-aware proxies for reaching non-convergent regions, but they do not necessarily correspond to distinct root-cause faults. Nearby crashes may reflect multiple manifestations of the same issue. The artifact includes an approximate clustering check of exported crash seeds, but clustering does not replace root-cause analysis or injected-fault benchmarks for measuring distinct failures. In addition, the use of ``concolic'' is intentionally modest: SYM-FUZZ reasons only over a small, discrete classical hyperparameter space and does not encode real-valued circuit parameters or quantum semantics.

\textbf{External validity and scalability.}
The evaluation uses two small workloads on a noiseless simulator, so the results may not generalize to larger HQC programs, calibrated noisy simulators, or hardware backends. Failure-guided fuzzing also does not reduce the underlying cost of running the hybrid program, which grows with qubit count, circuit depth, and shots; it reallocates a fixed budget toward configurations that appear failure-prone. Scaling to larger ans\"atze and richer hyperparameter spaces will require stronger seed-selection policies, bounding heuristics, and more expressive symbolic constraints.

\section{Conclusions and Future Work}
\label{subsec:future-work}

This paper studied failure-guided fuzzing for HQC programs on two NISQ workloads: a 2-qubit VQE instance and a 4-qubit QAOA MaxCut instance. The study modeled hybrid inputs as classical optimizer configurations coupled with quantum circuit parameters, defined a non-convergence oracle, and compared five budgeted strategies: RH, ENUM, RAND-FUZZ, ENUM-FUZZ, and SYM-FUZZ. The results show that reusing crash information through local fuzzing is the primary source of improvement
over RH and ENUM, while concolic seed discovery in SYM-FUZZ provides additional benefits on VQE but is less stable on QAOA. These findings position failure-guided fuzzing as a complementary direction to existing quantum fuzzing and concolic-testing work by operating at the hybrid-driver level and treating $(c,\boldsymbol{\theta})$ jointly.

Future work should extend the prototype with richer symbolic reasoning, broader HQC benchmarks, calibrated noisy backends, and more diverse oracles and coverage metrics. A larger-scale analysis of failure diversity and root causes are also needed to move from raw crash counts toward more actionable testing feedback for reliable quantum software engineering.

\section*{Data Availability}

The artifact is available at
\href{https://doi.org/10.5281/zenodo.19718515}{doi:10.5281/zenodo.19718515}.

\bibliographystyle{IEEEtran}
\bibliography{refs}

\end{document}